\documentclass[11pt]{article}

\title{Difference-in-Differences with Spatial Spillovers\thanks{I am grateful to Taylor Jaworski, Adam McCloskey, Brantly Callaway, Pedro H.C. Sant'Anna, Damian Clarke, Daniel Kaffine, Tania Barham, Alexander Bentz, James Flynn, Brach Champion, Hannah Denker, the CU Boulder Labor Economics Seminar, the Urban Economics Association Conference, the WEAI Conference, and the University of Rochester HEEL Lab for helpful comments.}}
\author{\href{https://kylebutts.com/}{Kyle Butts}\thanks{University of Colorado Boulder, Economics Department (\href{mailto:kyle.butts@colorado.edu}{kyle.butts@colorado.edu})}}
\date{\today}

\usepackage[margin=1in]{geometry}

\usepackage{amsmath}
\usepackage{amsfonts}
\usepackage{amsthm}
\usepackage{graphicx}
\usepackage[shortlabels]{enumitem}
\usepackage{bm}

\usepackage[T1]{fontenc}
\usepackage[bitstream-charter]{mathdesign}

\usepackage{microtype}

\usepackage{tikz}

\usepackage{xcolor}

\definecolor{dark-maroon}{HTML}{5D0F0D}

\definecolor{purple}{HTML}{5601A4}
\definecolor{navy}{HTML}{0D3D56}
\definecolor{ruby}{HTML}{9a2515}
\definecolor{alice}{HTML}{107895}
\definecolor{daisy}{HTML}{EBC944}
\definecolor{coral}{HTML}{F26D21}
\definecolor{kelly}{HTML}{829356}
\definecolor{cranberry}{HTML}{E64173}
\definecolor{jet}{HTML}{131516}
\definecolor{asher}{HTML}{555F61}
\definecolor{slate}{HTML}{314F4F}

\definecolor{kylemagenta}{HTML}{c24074}
\definecolor{kyleblue}{HTML}{3b587a}

\usepackage{hyperref}
\hypersetup{
    colorlinks= true,
    allcolors = purple
}

\usepackage{natbib}
\setcitestyle{comma,aysep={}}

\newtheoremstyle{spacing}
{}
{}
{}
{}
{\bfseries}
{.\ }
{2.5mm}
{}

\theoremstyle{spacing}

\newtheorem{proposition}{Proposition}[section]

\newtheorem{assumption}{Assumption}
\newtheorem{remark}{Remark}[section]

\global\long\def\expec#1{\mathbb{E}\left[#1\right]}%
\newcommand{\condexpec}[2]{\mathbb{E}\left[#1 \ \vert \ #2\right]}
\global\long\def\one{\mathbf{1}}%

\usepackage{titling}
\usepackage{setspace}

\pretitle{\begin{spacing}{1}\begin{flushleft}\huge}
\posttitle{\end{flushleft}\end{spacing}\vspace{-5mm}}
\preauthor{\begin{flushleft}\LARGE}
\postauthor{\end{flushleft}\vspace{-7.5mm}}
\predate{\begin{flushleft}\Large\color{asher}}
\postdate{\end{flushleft}\vspace{-5mm}}

\renewenvironment{abstract}
 {\noindent\rule{\linewidth}{.5pt}\noindent}
 {\noindent\rule{\linewidth}{.5pt}}

\usepackage[explicit]{titlesec}

\titleformat{\section}
  {\large \bf }
  {\thesection \,---}
  {0.25em}
  {#1}
  
\titleformat{\subsection}
  {\fontsize{11}{10}\it}
  {\thesubsection.}
  {1em}
  {#1}

\let\oldfootnote\footnote
\renewcommand\footnote[1]{\oldfootnote{\ #1}}

\makeatletter
\renewcommand\@makefntext[1]{%
    \parindent 1em \noindent
    \hb@xt@1.8em{\hss\normalfont\@thefnmark.\hfill}#1
  }
\makeatother

\usepackage{enumitem}
\setitemize{labelindent=0.5em,labelsep=0.25cm,leftmargin=*}
\setenumerate{labelindent=0.5em,labelsep=0.25cm,leftmargin=*}

\usepackage{caption}

\DeclareCaptionLabelSeparator{threedash}{\,---\,}
\DeclareCaptionFont{jet}{\color{jet}}
\usepackage{subcaption}
\makeatletter
\let\input\@@input
\makeatother

\usepackage{adjustbox}

\usepackage{array}
\renewcommand\arraystretch{1.25}

\usepackage[flushleft]{threeparttable}
\usepackage{booktabs}

\usepackage{tabularx}
\newcolumntype{L}{>{\raggedright\arraybackslash}X}
\newcolumntype{R}{>{\raggedleft\arraybackslash}X}
\newcolumntype{C}{>{\centering\arraybackslash}X}

\usepackage{pdflscape}

\usepackage{fancyhdr}
\fancypagestyle{lscaped}{%
    \fancyhf{}
    
    \textnormal
    \fancyfoot{%
        \tikz[remember picture,overlay]
        \node[outer sep=2.5cm,above,rotate=90] at (current page.east) {\thepage};
    }
}

\usepackage{pgfplots}
\pgfplotsset{compat=newest}
\usepgfplotslibrary{groupplots}
\usepgfplotslibrary{polar}
\usepgfplotslibrary{smithchart}
\usepgfplotslibrary{statistics}
\usepgfplotslibrary{dateplot}
\usepgfplotslibrary{ternary}
\usetikzlibrary{arrows.meta}
\usetikzlibrary{backgrounds}
\usepgfplotslibrary{patchplots}
\usepgfplotslibrary{fillbetween}
\pgfplotsset{%
    layers/standard/.define layer set={%
        background,axis background,axis grid,axis ticks,axis lines,axis tick labels,pre main,main,axis descriptions,axis foreground%
    }{
        grid style={/pgfplots/on layer=axis grid},%
        tick style={/pgfplots/on layer=axis ticks},%
        axis line style={/pgfplots/on layer=axis lines},%
        label style={/pgfplots/on layer=axis descriptions},%
        legend style={/pgfplots/on layer=axis descriptions},%
        title style={/pgfplots/on layer=axis descriptions},%
        colorbar style={/pgfplots/on layer=axis descriptions},%
        ticklabel style={/pgfplots/on layer=axis tick labels},%
        axis background@ style={/pgfplots/on layer=axis background},%
        3d box foreground style={/pgfplots/on layer=axis foreground},%
    },
}

\hypersetup{
    pdftitle={Difference-in-Differences Estimation with Spatial Spillovers},    pdfauthor={Kyle Butts}
}

\usepackage{etoolbox}

\newtoggle{INCLUDECOMMENTS}
\toggletrue{INCLUDECOMMENTS}

\begin{document}

\begin{titlepage}
    \maketitle

    \begin{abstract}
        {\small
        Empirical work often uses treatment assigned following geographic boundaries. When the effects of treatment cross over borders, classical difference-in-differences estimation produces biased estimates for the average treatment effect. In this paper, I introduce a potential outcomes framework to model spillover effects and decompose the estimate's bias in two parts: (1) the control group no longer identifies the counterfactual trend because their outcomes are affected by treatment and (2) changes in treated units' outcomes reflect the effect of their own treatment status and the effect from the treatment status of ``close'' units. I propose conditions for non-parametric identification that can remove both sources of bias and semi-parametrically estimate the spillover effects themselves including in settings with staggered treatment timing. To highlight the importance of spillover effects, I revisit analyses of three place-based interventions.
    
        \par~\par\noindent
        \noindent{\color{asher} JEL Classification Number:} C01, C21, C23, R15, R58
        \par
        \noindent{\color{asher} Keywords:} difference-in-differences, spatial econometrics, spillover effects, causal inference, place-based interventions
        \par\vspace{-2.5mm}
        }
    \end{abstract}
\end{titlepage}

\section{Introduction}

Empirical work in economics often considers settings where a policy is targeted to groups of units defined by geographic boundaries, but the effect of treatment spills over onto `nearby' units.\footnote{The framework of this paper applies to any setting with a well-defined measures of distance, e.g. geographic distance, economic distance such as supply chains, node distance in a graph, or social relationships in schools or cities.} For example, individuals in control areas can travel to the treated areas and receive treatment (e.g. a hospital opening serves nearby residents), or shocks to a labor market can affect nearby areas (e.g. a new factory opening increases service sector spending in the entire commuting zone). In some cases, the treatment effect may depend on the treatment status of other units. The concentration of treatment can increase its impact through knowledge-sharing between firms, while congestion or competition may decrease its effects, such as when wage bids mitigate employment effects of factory openings.

Estimation of treatment effects in these settings typically proceed in two ways. One approach involves using a structural model to account for general equilibrium effects. For example, trade models generate a market access term that effectively controls for such effects \citep{Donaldson_Hornbeck_2016} or the common Manski linear-in-means model controls for the average characteristics of a unit's peers \citep{manski1993identification,goldsmith2013social}. Another strategy involves excluding units located close to the treatment area (or including a dummy variable) to avoid any spillover biases on the effects, which is preferred by researchers who wish to remain agnostic to structural assumptions. This paper adopts the latter non-parametric approach and formalizes the identification assumptions required to interpret the estimates causally. 

To do so, this paper introduces a potential outcomes framework following \citet{vazquez2022identification} and extends the framework to difference-in-differences settings. Since potential outcomes are now characterized by a unit's treatment status as well as the treatment status of other units, there are many potential treatment effects that can be formualted. I study two relevant treatment effects. The second effect of interest is the `total effect' which is the relevant parameter for national policymakers who want to answer ``What is the average effect of implementing the entire treatment regime?''. This treatment effect summarizes the post-hoc effects of going from the world with no treatment to the observed treatment vector. First, there is the `switching effect' which is the relevant parameter for local policymakers who want to answer ``What is the effect of switching my treatment status, holding fixed all other units' treatment?''.

After defining these effects, I discuss non-parametric identification results extending standard parallel-trends assumptions to settings with spillovers. First, I demonstrate that the `total effect' can be identified under assumptions that spillovers are ``local'' and only affect nearby units, while parallel trends exist between treated and non-affected ``far-away'' units. This would hold in the case where individuals can travel to access treatment and therefore spillovers are limited by travel costs. However, this assumption may not hold in the case of large general equilibrium shocks that affect units far away from the treatment area, such as when New York's economy impacts San Francisco.

Estimating the switching effect without parametric assumptions on spillovers is challenging since the effect's magnitude often depends on the treatment status of neighboring units. For instance, the construction of low-cost medical facilities will have more substantial treatment effects when neighboring areas lack such facilities. As a result, switching effects must be defined separately for each level of exposure, and estimating the effect at a specific level necessitates knowledge of which control units share the same level of exposure. I demonstrate that identification requires \emph{either} parametrization of spillovers \emph{or} certain constraints on the homogeneity of spillover effects.

Subsequently, the commonly used difference-in-differences estimator is evaluated. The study demonstrates that this estimator produces biased estimates for both the total effects. This bias arises from the fact that untreated units that are "close" to treated units experience treatment effects, thereby failing to identify the counterfactual trend. The difference-in-differences estimate averages the spillover onto the "close" control units into the untreated units' change in outcomes. As a result, the spillover is subtracted from the estimated treatment effect, introducing bias in the opposite sign of the spillover effect. 

Then, I formalize the identification strategy commonly used to adjust for spillovers by including an indicator variable for units affected by spillovers (or removing them from the sample) in the two-way fixed-effects model. I show, under mild assumptions, that this method will remove \emph{all bias} from the treatment effect estimate so long as the indicator captures all units affected by spillovers. However, adding this indicator changes the underlying control group to the control units not included in the indicator variable and hence a researcher must justify why this subsample will form a valid control group for the parallel trends assumption. Importantly, my method does not require researchers to make any assumptions about how spillover effects propagate across space. The primary assumption required is selecting the maximum distance from treated units that spillovers can occur.\footnote{Since treatment effects typically decay over distance, this is not a strong assumption. Even if the maximum distance is not large enough, most of the bias will still be removed. On the other hand, making the maximum distance too large will increase the variance in the estimates as there are fewer remaining control units to identify the parallel counterfactual unit.} 

The modified difference-in-differences framework with an added spillover indicator provides a poor estimate of spillover effects. The estimate averages spillover effects over all units with the indicator equal to 1 which can include units that are unaffected by spillovers. Therefore the estimate will be attenuated to zero. The above method can be improved by creating a set of distance bins from the treated units (e.g. being 0-20 miles, 20-40 miles, 40-60 miles from the treated units) and interacting them with a treatment indicator. Since a set of `rings' is collinear with a single indicator covering the bins, this estimator consistently estimates the treatment effect of interest. This estimator has the added benefit that each indicator will estimate the average spillover effect on treated/control units within that distance range which provides a more complete picture of which units are affected by treatment and at what magnitude. However, interpreting these estimates causally requires that each ring satisfies a parallel trends assumption with the `far-away' control units.\footnote{The choice of the distance bins depends on the economic context and in particular the source of the spillovers. In some cases, the bins can be determined in a data-driven way \citep{Butts_2021}.}

To show the importance of considering spillover effects and the utility of my estimators, I revisit analyses of place-based policies in urban economics in Section \ref{sec:tva}. I revisit the analysis of the Tennessee Valley Authority by \citet{Kline_Moretti_2014}. The Tennessee Valley Authority was a large-scale New Deal program that lowered the cost of power for industrial firms \citep{Kitchens_2014}. The scale of federal investment in the region was large and the pro-manufacturing benefits likely spread further than the Authority's boundary due to the electrification infrastructure and agglomeration economies \citep{Severnini_2014}. I show that estimation by difference-in-differences fails to account for these spillovers and therefore the authors obtain biased estimates of the total effect of the Tennessee Valley Authority. For agriculture employment, I find that the long-run spillovers cause the original estimates to be about 40 percent too small for agriculture employment and 40 percent too large for manufacturing employment.

Following this empirical application, I discuss how my framework fits into a larger discussion on identification strategies with place-based policies. I revisit conflicting results about United States' federal Empowerment Zones, a program that creates incentives for businesses to locate in high-poverty neighborhoods. States apply for Empowerment Zone status for areas that meet a list of requirements but due to limited funding, not all proposed areas were accepted. \citet{Busso_Gregory_Kline_2013} use Census Tracts from qualified but ultimately rejected applications that are typically far away from accepted Zones as a comparison group and find significant reductions in poverty. This empirical strategy is based on the idea that since they also qualified for the program, these comparison units would be on parallel trends without being contaminated by spillover effects from treatment. \citet{Neumark_Kolko_2010}, on the other hand, use census tracts within 1,000 feet of the Zone and find no statistically significant effects. This empirical strategy assumes that the borders are drawn somewhat randomly and therefore units being just outside will have parallel counterfactual trends. Although, the trade-off is that using close units is potentially problematic as they can experience spillover effects from the Zones. My framework can explain the differences in findings if there are positive spillover effects onto census tracts just outside Empowerment Zones. In appendix \ref{sec:oz}, I show an example of the importance in considering spillovers when chosing identification strategy by analyzing the effects of the 2017 Opportunity Zone place-based policy.

Last, in Section \ref{sec:event_study}, I extend estimation of treatment and spillover effects into the TWFE/event-study framework by extending the work of \citet{Gardner_2021}. This allows for a very common setting in which treatment turns on for different units in different periods. Their proposed two-stage estimator first estimates group and time fixed-effects using untreated/not-yet-treated observations. Since some control/not-yet-treated units can be affected by spillovers, these units must be removed to consistently estimate the group and time fixed-effects. Then these estimated unit and time fixed-effects are subtracted from the observed outcome \emph{in the full sample}. The resulting residuals are then regressed on the treatment and spillover variables to estimate treatment and spillover effects. In the appendix, I demonstrate the method by revisiting the analysis by \citet{Bailey_Goodman_Bacon_2015} of Community Health Centers which provided low-cost primary care to impoverished areas.

\subsection{Relation to the Literature} 

I contribute to the literature that focuses on estimation of treatment effects with spatial spillovers using the difference-in-differences framework in three ways \citep{Clarke_2017,Berg_Streitz_2019,Verbitsky-Savitz_Raudenbush_2012,Delgado_Florax_2015}. My paper is the first to consider non-parametric identification in terms of general potential outcomes. Other papers derive results for specific spillover mappings. If I assume the particular functional forms for potential outcomes, I arrive at the same bias equation as theirs. The literature has focused on estimating the total effect, so I further contribute by discussing alternative treatment effect estimates and identification results. Last, my paper also advances the literature by considering estimation of treatment effects and spillover effects in the event-study framework which allows for the common occurrence of staggered treatment adoption.

This paper relates to a broad literature on spillover effect estimation in randomized experiments.\footnote{See \citet{Sobel_2006} and \citet{Hudgens_Halloran_2008} for early work. See \citet{Hu_Li_Wager_2021}, \citet{Savje_Aronow_Hudgens_2021}, and \citet{vazquez2022identification} and references therein for recent work on this.} There are two main strands of this literature. First, there is a large literature on the estimation of treatment effects in the presence of spillovers using a `partial identification' framework where units are in distinct treatment clusters and outcomes depend on the treatment status within the observation's cluster only.\footnote{ \citet{Angelucci_DiMaro_2016} provides an overview of estimation of treatment effects in the presence of ``within-group'' spillovers. Examples in the literature include: \citet{Halloran_Struchiner_1995} consider community-vaccine rates in epidemiology; \citet{Miguel_Kremer_2004} consider deworming programs in Kenyan schools; \citet{Sobel_2006} considers interference in the Moving to Opportunity Program; and \citet{Angrist_2014} studies the context of school peer effects.} Estimation compares units in the partially treated clusters with control units in completely untreated clusters which do not receive spillover effects. This allows standard difference-in-differences estimation of both the total effect (treatment effect on the treated) and spillover effects (treatment effect on the untreated in the treated clusters). However, in many settings distinct clusters are not plausible, e.g. counties near the border of a state can affect each other. My proposed estimator generalizes this method by using control units that are unaffected by treatment without requiring a notion of clusters.

There is also a nascent literature exploring estimation of treatment and spillover effects which does not require a completely untreated cluster in experimental settings (\citet{Savje_Aronow_Hudgens_2021}; \citet{vazquez2022identification}; \citet{Hu_Li_Wager_2021}; \citet{yu2022estimating}). Those papers' identification results rely on \emph{design-based} assumptions around the treatment-assignment mechanism, while this paper relies on \emph{model-based} assumptions, based on a modified parallel-trends assumption, for identification in non-experimental settings. I contribute to this literature by formalizing identification results of different treatment effects in non-experimental settings. 

The rest of the paper is structured as follows. Section \ref{sec:po_framework} presents the potential outcomes framework, defines the estimand of interest, discusses identification of effects, and shows the resulting bias from estimating a classical difference-in-differences model. Section \ref{sec:estimation} focuses on the estimation of the total effect and spillover effects. Section \ref{sec:tva} presents an application for evaluating the Tennessee Valley Authority program. Last, Section \ref{sec:event_study} discusses briefly how to incorporate spillovers into event-study estimation following the insights from \citet{Gardner_2021} and \citet{Borusyak_Jaravel_Spiess_2021}. 

\section{Potential Outcomes Framework}
\label{sec:po_framework}

This section presents a potential outcome framework that formalizes spatial spillovers. The goal of this exercise is to define possible treatment effects of interest since they are no longer uniquely defined in settings with spillovers. Following the canonical difference-in-differences framework, I observe a panel of units for periods $t \in \{0,1\}$ and treatment turns on between periods for some of the units.\footnote{This can be extended to multiple periods by replacing $Y_{i0}$ and $Y_{i1}$ throughout with pre- and post- sample averages of $Y_{it}$. This framework is extended to staggered treatment adoption in Section \ref{sec:event_study}.} Let $D_{i}$ be an indicator equal to one if unit $i$ is treated and therefore treatment status is defined as $D_{it} = D_{i} * \one_{t = 1}$.

Potential outcomes, denoted $Y_{it}(D_i, h_i(\vec{D}))$, are a function of unit $i$'s own treatment-status $D_i$ and, departing from the standard potential outcomes framework, of a function of the entire vector of treatment assignments $h_i(\vec{D})$ where $\vec{D} \in \{0,1 \}^n$ denotes the $n$-dimensional vector of all unit treatments. The non-negative scalar- or vector-valued function $h_i(\vec{D})$ is referred to as an `exposure mapping'. This setup follows \citet{vazquez2022identification} who studies identification of spillover effects under an experimental setting. Assumption (\ref{assumption:random}) allows me to consider observations as coming from a larger population (sometimes called a superpopulation).\footnote{Note that throughout, expectations and inference are conditional on the treatment vector (and hence exposure mappings) as in \citet{Borusyak_Jaravel_Spiess_2021}. In design-based settings, the distribution of exposure mappings is known so expectations can be taken without conditioning on the vector of exposure mappings. See \citet{Savje_Aronow_Hudgens_2021} and \citet{xu2023difference} for discussion of estimation and inference when the treatment design is known.} 

\begin{assumption}[Random Sampling]\label{assumption:random}
  $\{ Y_{i0}, Y_{i1} \}_{i = 1}^{n}$ is an independent and identically distributed sample of panel data.
\end{assumption}

The exposure mapping formalizes violations to the stable-unit treatment value assumption (SUTVA) by summarizing the intensity at which unit $i$ is affected by spillovers. Without any restrictions on the exposure mapping, nonparametric identification of treatment effects is impossible. The main assumption for identification, as is formalized in section \ref{sec:remove_bias}, is that when unit $i$ is sufficiently `far' away, it has no exposure to spatial spillovers and $h_i(\vec{D}) = \vec{0}$ . In the world of no spillovers, that is $h_i(\vec{D}) = \vec{0}$ for all individuals, potential outcomes would return to only depending on own-treatment status.\footnote{For example, $h_i(\vec{D})$ could be an indicator for being within a certain distance of a treated unit. Additionally, exposure could be defined as a spatial decay function where exposure is declining in the distance of treated units.}

Often, researchers will try to parameterize potential outcomes by reasoning through the nature of the spillovers and subsequent estimation will rely on the functional form assumptions made by researchers. This paper will take an alternative approach to identification and estimation which will allow for estimation of treatment effects without requiring direct knowledge of the functional form of the exposure mapping/potential outcomes. 

In particular, I will redefine the standard difference-in-differences identifying assumptions in the context of the new potential outcome framework. First, I assume that units do not adjust their actions in period $0$ from knowledge of future treatment:
\begin{assumption}[No Anticipation]\label{assumption:no-anticipation}
    $Y_{i0}(D, h) = Y_{i0}(0,0)$ for all values of $D$ and $h$.
\end{assumption}
Second, I formalize the parallel counterfactual trends assumption as follows:
\begin{assumption}[Parallel Counterfactual Trends]\label{assumption:parallel}
    Counterfactual trends do not depend on $D_i$:
    \[ 
        \expec{Y_{i1}(0, \vec{0}) - Y_{i0}(0, \vec{0}) \ \vert \ D_i = 1 } = 
        \expec{Y_{i1}(0, \vec{0}) - Y_{i0}(0, \vec{0}) \ \vert \ D_i = 0 }
    \]
\end{assumption}
This assumption states that in the absence of treatment and with zero exposure (not just the absence of individual $i$'s treatment), the change in potential outcomes from period 0 to 1 do not depend on treatment status. This generalizes to the classic parallel counterfactual trends when SUTVA is satisfied because then every unit has zero exposure. 

\subsection{Treatment and Spillover Effects}

In setups without spillovers, the treatment effect for an individual unit is well defined as $\tau_i \equiv Y_{i1}(1) - Y_{i1}(0)$, i.e. the effect of turning on treatment for a unit, and the above two assumptions would lead to the identification of the average effect by the traditional difference-in-differences estimator. In the presence of spillovers, multiple treatment effects can be defined in this setting. This subsection will define them and discuss their relative advantages/disadvantages. 

The natural analogue to the above treatment effect which I will label the `switching effect' is:
\[
    \tau_{i, switch}(\vec{h}) \equiv Y_{i1}(1, \vec{h}) - Y_{i1}(0, \vec{h}).
\] 
This is the effect of changing only unit $i$'s treatment status while keeping their exposure to spillovers constant at some value $\vec{h}$. This treatment effect is policy-relevant as it summarizes what would happen if a `local' policymakers decide to enact the policy for unit $i$ (implicity keeping $\vec{h}$ constant).\footnote{This is what \citet{Savje_Aronow_Hudgens_2021} call the `assignment-conditional unit-level treatment effect'.} It is important to note that the switching effect can depend on the level of exposure. For instance, consider the construction of libraries in towns. The effect of library construction in a town that is far away from any other library (exposure of $\vec{0}$) can be different than a town that is very close to a neighboring town's library (large $\vec{h}$), hence the dependence of $\tau_{i, switch}$ on $\vec{h}$. Since the size of the switching effect depends on a unit's exposure level, we will consider an average switching effect at each value of exposure, 
\[
    \tau_{\text{switch}}(\vec{h}) \equiv \condexpec{Y_{i1}(1, h_i(\vec{D})) - Y_{i1}(0, h_i(\vec{D}))}{D_i = 1, h_i(\vec{D}) = \vec{h}}.
\]

The second policy-relevant parameter is the `total effect of treatment on the treated': 
\[
    \tau_{\text{total}} \equiv \expec{Y_{i1}(1, h_i(\vec{D})) - Y_{i1}(0, \vec{0}) \ \vert \ D_i = 1}.
\] 
As opposed to the switching effect, which keeps exposure constant, the total effect looks at turning on treatment \emph{and} going from zero exposure to $h_i(\vec{D})$ simultaneously. This can be thought of as going from the world with a \emph{complete} absence of treatment, $\vec{0}$ to the current treatment vector $\vec{D}$. This treatment effect is helpful for `national' policymakers to evaluate what \emph{were} the effects of the entire vector of \emph{enacted policies}.\footnote{\citet{yu2022estimating} label the `total effect' to be $\expec{Y_{i1}(1, h_i(\vec{D})) - Y_{i1}(0, \vec{0})}$ where the average is over all units including control units. Identification of this effect relies on assumptions about experimental-design so this paper does not pursuit identification of this effect.}

To better understand how the above treatment effects are connected, it will be useful to formalize a unit's `spillover effect' as:
\[
    \tau_{i, \text{spill}}(D_i, h_i(\vec{D})) \equiv Y_{i1}(D_i, h_i(\vec{D})) - Y_{i1}(D_i, \vec{0}).
\] 
The spillover effect measures the difference in potential outcomes between being exposed at intensity $h_i(\vec{D})$ and not being exposed. This effect can differ for either treated or control units as the magnitude or even the nature of spillovers might differ between treated and untreated units. That is $\tau_{i, \text{spill}}(1, h_i(\vec{D}))$ does not necessarily equal $\tau_{i, \text{spill}}(0, h_i(\vec{D}))$. We can then average over treated/control units to form average spillover effects: 
\[
    \tau_{\text{spill}}(D) \equiv \expec{Y_{i1}(D, h_i(\vec{D})) - Y_{i1}(D, \vec{0}) \ \vert \ D_i = D},
\]
where the expectation is over individuals and hence over different exposures. This will give us the average spillover effect among \emph{all} treated/control units regardless if they experience any spillover effects. It is often also of interest to estimate the average spillover effects on subsets of units (e.g. all units experiencing non-zero exposure). Estimation of spillover effects will be discussed in Section \ref{sec:spill}.

With spillover effects formally defined, the difference between the proposed estimands become more clear by the following algebraic identity:
\begin{align*}
    &\overbrace{Y_{i1}(1, h_i(\vec{D})) - Y_{i1}(0, h_i(\vec{D}))}^{\text{Switching Effect}} = \\
    &\quad\quad \underbrace{\underbrace{Y_{i1}(1, 0) - Y_{i1}(0,0)}_{\text{Direct Effect}} \ + \ \underbrace{Y_{i1}(1, h_i(\vec{D})) - Y_{i1}(1, 0)}_{\text{Spillover on Treated}}}_{\text{Total Effect on Treated}} \ - \ \underbrace{(Y_{i1}(0, h_i(\vec{D})) - Y_{i1}(0,0))}_{\text{Spillover on Control}} 
\end{align*}

From this identify, we see that the switching effect can be decomposed into two parts. First, the unit is affected by changing its treatment status as summarized by the `direct effect'. Second, the unit changes from experiencing spillover effects as a control unit to experiencing spillover effects as a treated unit. Additionally, the above work shows that the switching effect is closely related to the total effect. The primary difference is that the total effect is concerned with post-hoc analysis so that the reference potential outcome is $Y_{i1}(0,0)$. On the other hand, the switching effect has a reference of $Y_{i1}(0, h_i(\vec{D}))$ so that when entering treatment, the spillover on control units is lost.

\subsection{What Does Difference-in-Differences Estimate?}

With treatment effects properly defined, I will first derive what the standard difference-in-differences estimand identifies under parallel trends assumption. 

\begin{proposition}[Decomposition of Difference-in-Differences Estimand]\label{thm:bias}\ \\    
    If assumptions (\ref{assumption:random}), (\ref{assumption:no-anticipation}) and \nameref{assumption:parallel} hold, the population difference-in-differences estimand can be decomposed as follows:
    \begin{align}
        &\underbrace{\expec{Y_{i1} - Y_{i0} \ \vert \ D_i = 1 } - \expec{Y_{i1} - Y_{i0} \ \vert \ D_i = 0 }}_{\text{Difference-in-Differences}} \label{eq:did} \\ 
        &\quad\quad= \tau_{\text{total}} - \tau_{\text{spill}}(0)
    \end{align}
\end{proposition}

The proof is given in Appendix \ref{sec:proofs}, but the intuition is as follows. The change in outcomes among control units combines the parallel counterfactual trend with the average spillover effect onto control units. Since $\hat{\tau}$ is found by subtracting the change in outcomes among the control units, we subtract the average spillover effect onto the control, $\tau_{\text{spill}}(0)$. 

Researchers typically estimate the canonical two-way fixed-effects model, 
\begin{equation}\label{eq:twfe}    
    y_{it} = \tau D_{it} + \mu_i + \lambda_t + \epsilon_{it}.
\end{equation}
The estimator $\hat{\tau}$, which is well understood to be the sample analog of (\ref{eq:did}), is, therefore, a biased estimate for $\tau_{\text{total}}$ in the presence of spillovers. In particular, if a researcher estimates a difference-in-differences model and reports $\hat{\tau}$ as an estimate of the total effect, their estimate would be biased in the presence of spillover effects on control units. If spillover effects are of the same sign as the total effect, then estimates would be attenuated towards zero. On the other hand, if spillovers are the opposite sign, then results would increase in magnitude. 

\subsection{Identification of Treatment Effects}\label{sec:remove_bias}

The following subsection will talk about the identification of treatment effects and present corresponding estimators for each effect. 

\subsubsection{Switching Effect}

It is difficult to identify the switching effect in general as identifying the unobserved counterfactual outcome $Y_{i1}(0, h_i(\vec{D}))$ requires knowledge of which control units have the same level of exposure. The reason for this can be seen by rewriting this in terms of a difference-in-differences style estimand:
\begin{align*}
    \tau_{\text{switch}}(\vec{h}) &= \mathbb{E} \left[ Y_{i1}(1, h_i(\vec{D})) - Y_{i0}(0,\vec{0}) \mid D_i = 1, h_i(\vec{D}) = \vec{h} \right] \\ 
    &\quad\quad - \mathbb{E} \left[ Y_{i1}(0, h_i(\vec{D})) - Y_{i0}(0,\vec{0}) \mid D_i = 1, h_i(\vec{D}) = \vec{h} \right].
\end{align*}
Writing it in this way highlights the two problems in estimating the second term. First, this requires knowledge of the unobserved exposure mapping to be correctly specified to identify control units with $h_i(\vec{D}) = \vec{h}$. More, this estimand requires a more demanding parallel trends assumption.\footnote{This problem is similar to those highlighted by \citet{Callaway_Goodman-Bacon_SantAnna_2021} for difference-in-differences with continuous treatment.}  

\begin{assumption}[Switching Effect Parallel Trends]\label{assumption:parallel-switching}
    For a given exposure, $\vec{h}$, counterfactual trends do not depend on $D_i$:
    \[ 
        \expec{Y_{i1}(0, \vec{h}) - Y_{i0}(0, \vec{0}) \ \vert \ D_i = 1, h_i(\vec{D}) = \vec{h} } = 
        \expec{Y_{i1}(0, \vec{h}) - Y_{i0}(0, \vec{0}) \ \vert \ D_i = 0, h_i(\vec{D}) = \vec{h} }
    \]
\end{assumption}

This assumption requires parallel trends among treated and control units at exposure $\vec{h}$. Under this assumption, the switching effect can be identified by a difference-in-differences estimand calculated using only units with an exposure of $\vec{h}$.

\begin{proposition}[Identification of the Switching Effect]
    For a given exposure, $\vec{h}$, if assumption (\ref{assumption:random}), (\ref{assumption:no-anticipation}), and \nameref{assumption:parallel-switching} holds, then 

    \begin{align*}
        \tau_{\text{switch}}(\vec{h}) &= \mathbb{E} \left[ Y_{i1}(1, h_i(\vec{D})) - Y_{i0}(0,\vec{0}) \mid D_i = 1, h_i(\vec{D}) = \vec{h} \right] \\ 
        &\quad\quad - \mathbb{E} \left[ Y_{i1}(0, h_i(\vec{D})) - Y_{i0}(0,\vec{0}) \mid D_i = 0, h_i(\vec{D}) = \vec{h} \right].
    \end{align*}
\end{proposition}

The switching effect can be estimated by a difference-in-differences estimator on the subsample of units with exposure $\vec{h}$. However, this approach relies on correct specification of the exposure mapping which is a difficult assumption to maintain in practice. Researchers may be tempted to introduce the parameterized exposure mapping linearly to ``compare individuals at the same level $\vec{h}$''. However, this assumes additive separability between the treatment dummy and the exposure mapping. Additive separability would imply the switching effect does not depend on exposure level, i.e. the switching effect is constant across $\vec{h}$.

\subsubsection{Total Effect}

To identify the total effect, additional assumptions on the nature of spillover effects are needed. In particular, I will formalize the idea that spillovers are `local' in that units are only affected by treatment if they are near treatment. For example, if treatment has to be accessed in person, then it is natural that further away places are not affected by treatment. However, this assumption may fail under general equilibrium shocks that do not necessarily decay over distance.

\begin{assumption}[Spillovers Are Local]\label{assumption:local}
    There exists a distance $\bar{d}$ such that 
    
    (i) For all units $i$,
    \[ 
        \min_{j: \ D_j = 1} d(i,j) > \bar{d} \implies h_i(\vec{D}) = \vec{0}. 
    \]

    (ii) There are treated and control units, $i$, such that $\min_{j: \ D_j = 1} d(i,j) > \bar{d}$.
\end{assumption}

Part (i) of assumption (\ref{assumption:local}) requires that spillovers are `local' in that units are no longer exposed to spillovers after some maximum distance $\bar{d}$. Part (ii) of Assumption (\ref{assumption:local}) requires that there exists control and treated units with no exposure. This assumption is far less strict than the identifying assumption for the switching effect in that we only need to identify units with non-zero exposure and don't need to parameterize the exposure mapping any further. For a given $\bar{d}$ that satisfies Assumption (\ref{assumption:local}), let $S_{i}$ be an indicator equal to one if $\min_{j: \ D_j = 1} d(i,j) \leq \bar{d}$, i.e. unit $i$ is within $\bar{d}$ miles of the closest treated unit. Similar to above, let $S_{it} = S_i * \one{t = 1}$ be the treatment indicator.

The intuition behind identifying the total effect requires that we use control units experiencing no spillover effects to identify the counterfactual trend. Since we are leveraging subgroups of control units, we need a parallel trends assumption for each subgroup:

\begin{assumption}[Total Effect Parallel Trends]\label{assumption:parallel-mod}
    Counterfactual trends do not depend on $D_i$ and $S_i$:
    \[
        \expec{Y_{i1}(0, \vec{0}) - Y_{i0}(0, \vec{0}) \ \vert \ D_i = 1}= 
        \expec{Y_{i1}(0, \vec{0}) - Y_{i0}(0, \vec{0}) \ \vert \ D_i = 0, S_i = 0}
    \]
\end{assumption}

It is worth remarking on \nameref{assumption:parallel-mod} in terms of applied work. Often, individuals use only a subsample of observations that are close to treated observations in estimating their difference-in-differences model. The idea is that close units are more likely to be on similar counterfactual trends than units that are further away. The identification result and its corresponding estimator are formed by using further away units (i.e. units with $S_i = 0$). In this case, if parallel trends only hold locally, then focusing on further away units can fix the spillover bias but worsten bias from non-parallel trends. With this assumption, we have the following identification result for the total effects:

\begin{proposition}[Identification of Total Effects]\label{thm:remove_bias}
    If assumption (\ref{assumption:random}), (\ref{assumption:no-anticipation}), 
    
    \noindent\nameref{assumption:local}, and \nameref{assumption:parallel-mod} hold, then (i) 
    \begin{equation}\label{eq:id_total}
        \expec{Y_{i1} - Y_{i0} \ \vert \ D_i = 1} - \expec{Y_{i1} - Y_{i0} \ \vert \ D_i = 0, S_{i} = 0} = \tau_{\text{total}}.
    \end{equation}
\end{proposition}

The proof is given in Appendix \ref{sec:proofs}, but the intuition is simple. Conditioning on $S_i = 0$ will use only observations with $h_i(\vec{D}) = 0$ which gives rise to the identification of the total effect. 
This proposition shows that a researcher does not need to know how the spillovers occur over space to have an unbiased estimate for treatment effects. All that is required to identify the total effect is being able to classify which units are considered `close' to treatment.\footnote{In contexts with different measures of `distance', note that this result only requires an indicator that captures all affected units. For example, thinking about spillovers between industries, this requires the researcher to correctly identify which industries are affected by spillovers and which are not.} 

\begin{remark}
  To highlight the importance of the modified parallel trends, we will compare the two estimands under the original \nameref{assumption:parallel}. The standard difference-in-differences estimate is biased from spillovers on nearby control units: 
  $$
    \expec{Y_{i1} - Y_{i0} \ \vert \ D_i = 1 } - \expec{Y_{i1} - Y_{i0} \ \vert \ D_i = 0 } = \tau_{\text{total}} - \tau_{\text{spill}}(0).
  $$
  On the other hand, using the $S_i = 0$ control group is biased by (potentially) non-parallel trends:
  \begin{align*}
    &\expec{Y_{i1} - Y_{i0} \ \vert \ D_i = 1} - \expec{Y_{i1} - Y_{i0} \ \vert \ D_i = 0, S_{i} = 0} \\
    &\quad\quad= \tau_{\text{total}} + (\expec{Y_{i1}(0, 0) - Y_{i0}(0,0) \ \mid \ D_i = 1} - \expec{Y_{i1}(0, 0) - Y_{i0}(0,0) \ \mid \ D_i = 0, S_i = 0}) 
  \end{align*}
  In this sense, there is a practical trade-off between two different sources of bias: spillover bias and the forming of potentially worse counterfactual conrol group.
\end{remark}

\section{Estimation of Effects}\label{sec:estimation}

\subsection{Estimation of Total Effect}

To estimate the total effect, we could replace the terms in proposition \ref{thm:remove_bias} with their sample analogs. However, it will prove advantageous to modify the two-way fixed-effects model to estimate treatment effects because we will extend it to estimate spillover effects in the following subsection. To estimate the total effect, one can run the regression specification (\ref{eq:twfe}) on the subsample of units with $D_i = 1$ or $S_i = 0$. From standard regression analysis the estimates of $\hat{\tau}$ are sample analogues of (\ref{eq:id_total}). 

Since $\hat{\tau}$ is a linear combination of sample averages, inference can be done using large-n asymptotics. However, it is important to account for two sources of correlation in the error structure. First, errors should be clustered by unit $i$ to allow for correlation across periods. Second, since assumption \ref{assumption:local} is predicated on the fact that nearby places affect one another, we should account for such spatial correlation by allowing for spatial correlations following \citet{Conley_1999}. More recent work by \citet{Ferman_2020} show that for large-n asymptotics to be used, the structure of spatial correlation must be limited in that errors are assumed to be uncorrelated after a certain cutoff distance. A natural candidate for this cutoff would be $\bar{d}$ used for the creation of $S_i$. 

A natural question is why wouldn't researchers make the distance large to guarantee they have an unbiased estimate of $\tau_{\text{total}}$? Equation (\ref{eq:id_total}) should make this problem clear. Since these estimates rely on averages among units with $S_i = 0$, as $\bar{d}$ increases, the number of units with $S_i = 0$ decreases which yields more variable estimates. On the other hand, having units that experience spillovers not included in the $S_i$ indicator will leave some bias in the estimate.\footnote{Although, spillover effects typically will grow weaker over distance and if distant units are mistakenly treated as if they have zero exposure, bias should be small.} Therefore there is a bias-variance trade-off in extending the extent of $S_i$ that should be balanced by researchers guided by their particular economic context. Additionally, each value of $\bar{d}$ corresponds to a different effective control group and hence a different parallel trends assumption. In separate work, \citet{Butts_2021} discusses data-driven methods to determine how far $S_i$ should be under a more strict parallel trends assumption.

\subsection{Estimation of Spillover Effects}\label{sec:spill}

This subsection turns to estimating averages of spillover effects themselves. To do so, we must extend the \nameref{assumption:parallel-mod} to allow include control units with $S_i = 1$. 
\begin{assumption}[Spillover Effect Parallel Trends]\label{assumption:parallel-spill}
  For $s = 0, 1$, we have: 
  \[
      \expec{Y_{i1}(0, \vec{0}) - Y_{i0}(0, \vec{0}) \ \vert \ D_i = 1} = 
      \expec{Y_{i1}(0, \vec{0}) - Y_{i0}(0, \vec{0}) \ \vert \ D_i = 0, S_i = s}
  \]
\end{assumption}

In this case, both treatment effects and spillover effects can be estimated with the following regression specification using the full sample:
\begin{equation}\label{eq:twfe_spill}
  Y_{it} = \tau D_{it} + \gamma_0 (1 - D_{it}) S_i + \mu_i + \lambda_t + \epsilon_{it}.
\end{equation}
Our total effect estimate, $\hat{\tau}$, from (\ref{eq:twfe_spill}) is consistent under \nameref{assumption:parallel-spill}. The spillover effect estimate $\hat{\gamma}_0$ in (\ref{eq:twfe_spill}) does not estimate the corresponding average spillover effects $\tau_{\text{spill}}(0)$. This is because $\hat{\gamma}_0$ average spillover effects across only untreated units with $S_i = 1$:
\[ 
    \expec{\hat{\gamma}_0} = \condexpec{\tau_{i, spill}(0)}{ S_i = 1, D_i = 0}.
\]
Since some units with $S_i = 1$ do not have positive exposure, this is not the average spillover effect for units with non-zero exposure either. For example, if only units really close to treatment receive spillover effects and $S_i$ contains a lot of units that experience no spillover effects, then $\hat{\tau}_{\text{spill, control}}$ could be estimated near zero even though some units experience substantial spillover effects.

It is possible to more accurately estimate spillover effects by using a set of concentric ring indicators (e.g 0-20 miles, 20-40 miles, and 40-60 miles from the nearest treated unit) instead of a single large ring indicator to more flexibly estimate the spillover effects. Equation (\ref{eq:twfe_spill}) can be modified as 
\begin{equation}\label{eq:twfe_rings}    
    y_{it} = \tau D_{it} + \sum_{j=1}^n (1-D_{it}) \text{Ring}_{ij} + \mu_i + \lambda_t + \epsilon_{it},
\end{equation}
where $\text{Ring}_{ij}$ is an indicator for unit $i$ being in ring $j$. However, this estimator relies on a yet more stringent parallel trends assumption that the untreated units in each ring are on the same trend as the untreated units with $S_i = 0$.\footnote{\citet{Butts_2021} takes this to the logical extreme where more and more rings are added. The parallel trends assumption becomes that untreated units are on the same trend at every distance from treatment.} Since the set of rings is collinear with a single large ring, the total effects estimate will continue to be identified. Each ring will estimate the average of $\tau_{i,spill}(0)$ for all observations within a given distance ring. Having estimates of average spillover effects at different distances allows a researcher to better understand the nature of the spillovers.

\begin{figure}[tb!]
    \caption{Comparison of Single vs. Multiple Rings Estimation of Spillover Effects}
    \label{fig:example_rings}

    \includegraphics[width = \textwidth]{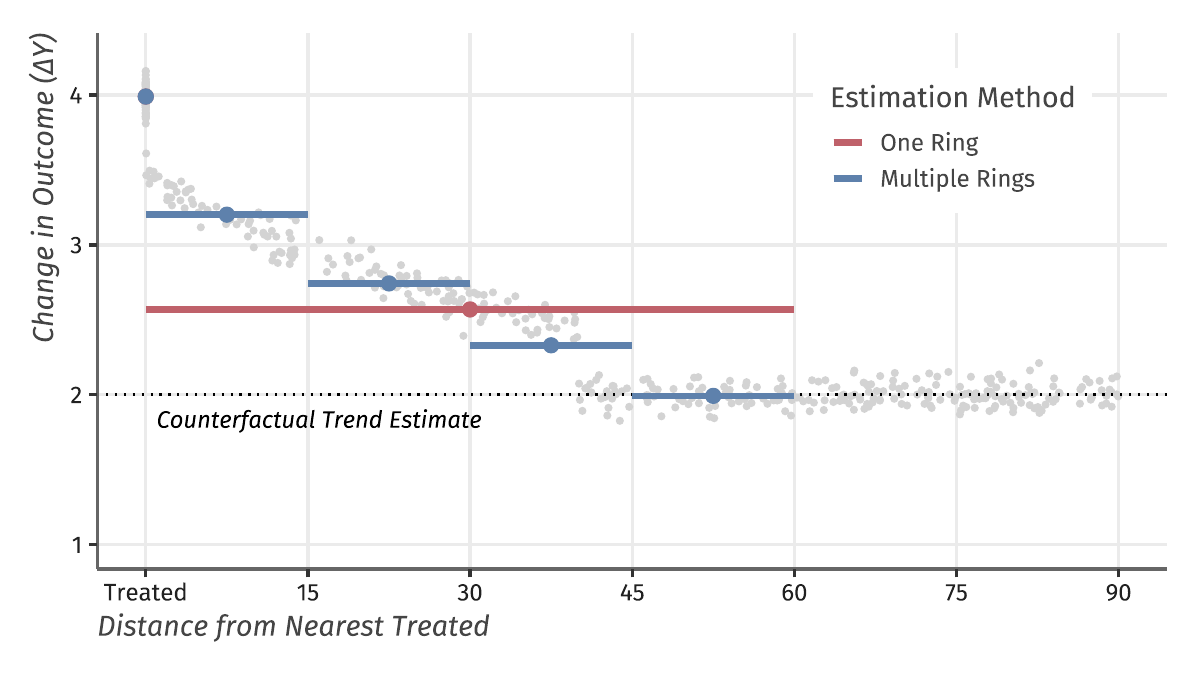}

    {\footnotesize \textit{Notes:} The figure compares estimates of equation (\ref{eq:twfe_rings}) with $S_i$ a ring of $(0, 60]$ miles from the nearest treated unit and of equation (\ref{eq:twfe_rings}) with rings for $(0,15], (15,30], (30,45], (45, 60]$ miles for a simulated set of data.}
\end{figure}

A simple example of how rings improve the estimation of spillover effects can be seen in Figure \ref{fig:example_rings}. The figure displays observations of $Y_{i,1} - Y_{i,0}$ on the y-axis and the distance to the nearest treated unit on the $x$-axis. Treated units are placed at a zero distance for visualization purposes. The example has a counterfactual trend of 2 so any difference from 2 is due to either treatment/spillover effects and the error term. 

The `One Ring' estimation method uses a single indicator for being between 0 and 60 miles from treatment. The difference between the bar and the estimated counterfactual trend is the estimated spillover effect which in this case is about 0.6. This number masks that units within the first 15 miles have a treatment effect of 1 or larger and units between 40 and 60 miles have 0 spillover effects. The `Multiple Rings' estimation method uses 4 indicators for every 15-mile interval (0-15, 15-30, 30-45, and 45-60 miles). This improves the estimate of a single ring in two ways. Multiple rings are better able to capture the decaying nature of spillover effects over distance. Furthermore, the ring that is included past the true maximum distance where spillover effects occur (45-60 miles) has a point estimate near zero. Note that this example works well because counterfactual trends are assumed to be constant across distance. If this weren't true, adding rings would require stronger and stronger parallel trends assumptions.

The rings method is a semi-parametric estimator in the sense that as the number of rings increase and their width shrinks, the estimates would trace out the spillover effect function (subject to $\bar{d}$ satisfying assumption (\ref{assumption:local})). However, the precision of estimates would decrease as the number of rings increases and therefore there is a bias-variance trade-off in the number and width of rings.\footnote{\citet{Clarke_2017} proposes a cross-validation technique for optimally choosing the number of rings and their techniques that uses the out-of-sample mean square prediction error as a way to estimate the bias-variance tradeoff. \citet{Butts_2021} provides a data-driven method to selecting rings under a more strict version of the parallel trends assumption.} 

One disadvantage of using a set of indicators for \textit{the nearest} treated unit is that some spillover effects are additive in the number of nearby treated units. In this case, summarizing exposure by the distance to the closest treated unit fails to capture important information on the location of other treated units and the coefficients on the ring indicators might be interpreted incorrectly. One potential solution is to count the number of treated units within each ring instead of creating indicators for the nearest treated unit. This additive version of rings however no longer removes all bias from the treatment effect estimate unless the exposure mapping is correctly specified. Therefore, if spillover effects are expected to be additive in nature, separate estimation of the total effect and spillover effects would be the preferred method. 

\section{Application in Place-Based Policy Analysis}
\label{sec:tva}

To illustrate the importance of accounting for spatial spillovers in the estimation of treatment effects, I revisit the analysis of the Tennessee Valley Authority (TVA) in \citet{Kline_Moretti_2014}. The TVA program was a large-scale federal investment started in 1934 that focused on the construction of dams and transportation canals in an attempt to modernize the Tennessee Valley's economy. By the end of WWII, the TVA became the largest single power supplier in the country and significantly lowered the cost of wholesale energy for factories.\footnote{More details on the program are found in \citet{Kline_Moretti_2014}. The effects on wholeseale electricity are discussed in \citet{Kitchens_2014}.} With over \$20 Billion (in 2000 dollars) spent which is hundreds of dollars transferred per person in the Authority, the impacts are very likely to extend past the authority's borders. \citet{Kline_Moretti_2014} analyze a range of outcome variables, but for succinctness, I use only (the log of) agricultural and manufacturing employment. Since the TVA primarily improved manufacturing industries through large-scale electrification, the authors predict that employment will grow in manufacturing and shrink in agricultural as workers switch to higher-paying manufacturing jobs. 

\begin{figure}[tb!]
    \caption{TVA Effective Sample and Spillover Variables}
    \label{fig:tva_sample}

    {\centering
        \resizebox{\textwidth}{!}{
            \includegraphics{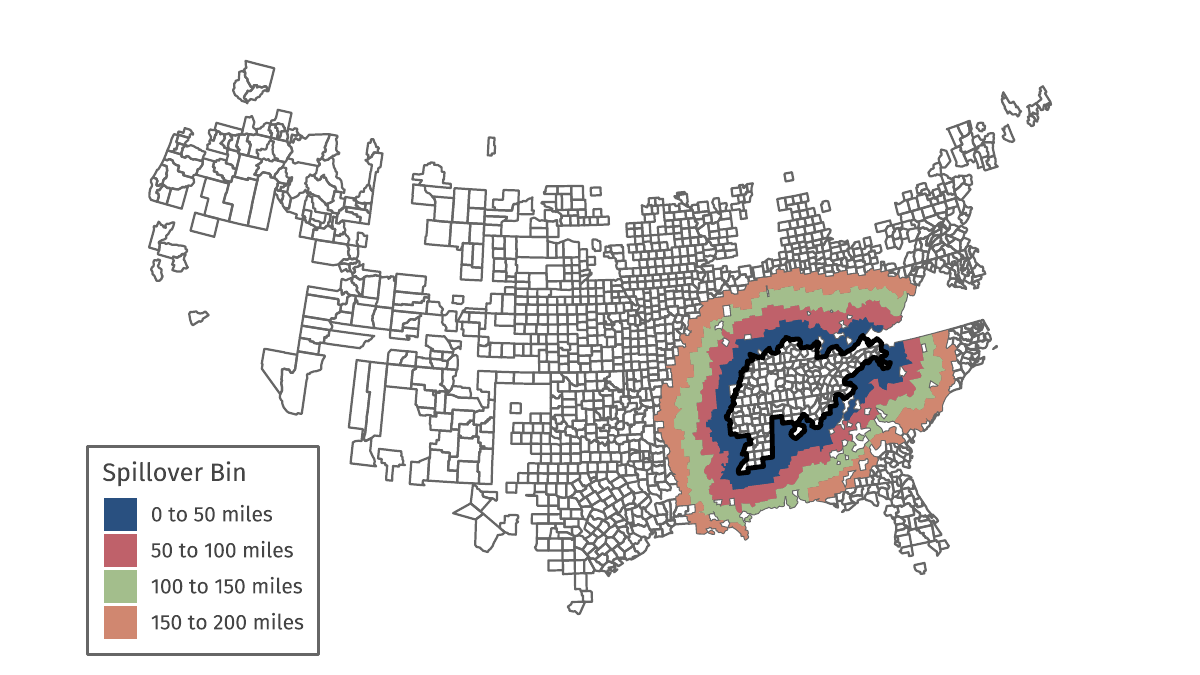}
        } 
    }
    {\footnotesize \textit{Notes:} The above figure plots all the counties used in the estimation. Counties that fall within the distance intervals $\{ (0, 50], (50, 100], (100, 150], (150, 200] \}$ measured in miles are colored by their respective bin.} 
\end{figure}

The analysis in \citet{Kline_Moretti_2014} begins by comparing changes in county-level outcomes from 1940 to either 1960 (short run effects) or 2000 (long run effects) between treated counties in the Authority and control counties outside. The primary specification is
\begin{equation}\label{eq:tva}
    y_{c, t} - y_{c, 1940} = \alpha + \text{TVA}_c \tau + X_{c, 1940} \gamma + (\varepsilon_{c, t} - \varepsilon_{c, 1940}),
\end{equation}
where $c$ denotes county, $\text{TVA}_c$ is an indicator variable for being in the Authority, and $y$ is a set of outcome variables (in logs).\footnote{The two-period difference-in-differences regression is equivalent to a first-difference regression. The authors use an Oaxaca-Blinder estimator on the first differences and the results of \citet{Kline_2011} show that this estimator is equivalent to a weighted difference-in-differences estimate. Their estimator does not differ much from the standard difference-in-differences results since the weights are not that different from uniform weights.} Pre-treatment control variables, $X_{c,1940}$, are interacted with $\text{Post}_t$ to allow for places to be on different long-term trends.\footnote{See footnote 8 in \citet{Kline_Moretti_2014} for a full listing of control variables.} To improve the likelihood of the parallel trends assumption, they run a logistic regression to predict being in the TVA based on their set of control variables $X_{c,1940}$ and keep only observations in the top 75\% of predicted probability. The counties used in the sample are presented in Figure \ref{fig:tva_sample}. \citet{Kline_Moretti_2014} estimate (\ref{eq:tva}) to identify the `local effect' of the TVA -- what I am calling the `total effect on the treated'. However, their point estimates compare, in part, changes in outcomes in TVA counties with changes in outcomes for neighboring counties that likely were impacted by the large-scale program. 

In the paper, the authors discuss the nature of spillovers that can occur. For agriculture employment, the authors claim that improved wages in the Authority will draw agriculture workers out of nearby counties. Hence they predict a negative spillover. For manufacturing, the sign is ambiguous. There could be positive spillovers if electrification brought cheap power and agglomeration economies to the neighboring areas. However, manufacturing could decline if firms chose to locate in the Authority that would have, in the absence of the program, decided to locate in nearby counties. My methodology will allow me to empirically test these predictions in the data and remove their bias from the treatment effect estimates. 

By comparing counties inside the Authority to counties on the other side of the border, the authors likely underestimate the negative effect on agricultural employment while the bias in the manufacturing effect is theoretically ambiguous. The authors do recognize the problem of these comparisons and remove counties that share borders with the authority, but due to the scale of the program, the spillovers are likely to extend further than this. Therefore, bias in their estimate will likely remain even after dropping contiguous counties. The estimation strategy I present keeps the observations near the TVA while controlling for spillover effects in a more rigorous manner.  

I extend their analysis to control for spatial spillovers in the difference-in-differences specification. To parametrize the exposure mapping, I use a set of rings as described in Section \ref{sec:spill}. Specifically, 
the specification with spillovers is given as follows:  
\begin{equation}\label{eq:tva_spillover}
    y_{i, t} - y_{i, 1940} = \alpha + \text{TVA}_i \tau + \sum_{j \in \text{Dist}} \text{Ring}_{ij} \delta_j + X_{i, 1940} \beta + (\varepsilon_{i, t} - \varepsilon_{i, 1940}),
\end{equation} 
where $\text{Dist} = \{(0, 50], (50, 100], (100, 150], (150, 200]\}$ measured in miles and define $\text{Ring}_{ij}$ as an indicator for being within the interval $d \in \text{Dist}$ away from the Authority and $t \in \{1960, 2000\}$. Figure \ref{fig:tva_sample} displays the four spillover variables by filling in each distance bin in a different color. The coefficients $\delta_j$ estimate the average spillover effect onto control units for each of these distance bins. From proposition \ref{thm:remove_bias}, $\hat{\tau}$ will be an unbiased estimate for the `total' effect so long as spillovers do not occur past 200 miles from the TVA.

The results of the long run analysis from 1940 to 2000 are presented in Panel A of Table \ref{tab:tva}. Each row contains the results for a different outcome variable (measured in logs). Columns (1)-(5) contain point estimates for $\tau$ and $\delta_j$'s in different specifications. The point estimates can be interpreted as decadal growth rates in outcomes. The column labeled difference-in-differences uses an ordinary least squares estimator for the specification without spillovers, equation (\ref{eq:tva}). This estimate finds a decline in agricultural employment of about $5.1\%$ per decade and an increase in manufacturing employment of about $5.6\%$ per decade. 

\begin{table}[!tb]
    \caption{Effects of Tennessee Valley Authority on Decadal Growth}
    \label{tab:tva}
    \renewcommand{\arraystretch}{1.1}

    \begin{adjustbox}{width = \textwidth, center}
        \begin{threeparttable}
            \begin{tabular}{@{} l c@{\extracolsep{20pt}}c@{\extracolsep{4pt}}cccc @{}}
                \toprule

                & \multicolumn{1}{c}{\textbf{Diff-in-Diff}} & \multicolumn{5}{c}{\textbf{Diff-in-Diff with Spillovers}} \\ 
                \cmidrule{2-2} \cmidrule{3-7}
                & & & TVA between & TVA between & TVA between & TVA between \\ 
                & TVA & TVA & 0-50 mi. & 50-100 mi. & 100-150 mi. & 150-200 mi. \\ 
                \textit{Dependent Var.} & (1) & (2) & (3) & (4) & (5) & (6) \\

                \toprule
                \multicolumn{7}{l}{\textbf{Panel A:} 1940-2000} \\
                \midrule
                
                 Agricultural employment     & $-0.0514^{***}$& $-0.0739^{***}$& $-0.0371^{***}$&    $-0.0164$   & $-0.0298^{***}$&  $-0.0157^{*}$ \\
                             &   $(0.0114)$   &   $(0.0142)$   &   $(0.0002)$   &   $(0.0114)$   &   $(0.0096)$   &   $(0.0088)$   \\
 Manufacturing employment    & $0.0560^{***}$ &    $0.0350$    & $-0.0203^{***}$&    $-0.0245$   &  $-0.0331^{*}$ & $-0.0296^{**}$ \\
                             &   $(0.0161)$   &   $(0.0218)$   &   $(0.0006)$   &   $(0.0281)$   &   $(0.0189)$   &   $(0.0142)$

                \\ \toprule
                \multicolumn{7}{l}{\textbf{Panel B:} 1940-1960} \\
                \midrule 

                 Agricultural employment     & $0.0940^{***}$ &  $0.0856^{*}$  &    $-0.0062$   &    $-0.0042$   &    $-0.0303$   &    $-0.0039$   \\
                             &   $(0.0275)$   &   $(0.0444)$   &   $(0.0474)$   &   $(0.0448)$   &   $(0.0403)$   &   $(0.0297)$   \\
 Manufacturing employment    & $0.0894^{***}$ &  $0.0993^{**}$ &    $0.0228$    &    $0.0225$    &    $-0.0055$   &    $-0.0066$   \\
                             &   $(0.0324)$   &   $(0.0432)$   &   $(0.0510)$   &   $(0.0561)$   &   $(0.0347)$   &   $(0.0253)$

                \\ \bottomrule
            \end{tabular}
            
            \begin{tablenotes}\footnotesize
                \item \textit{Notes.} Each row corresponds to an outcome variable. Each cell is the point estimate and the standard error for the variable described in the column title. All standard errors are Conley standard errors with a correlation cutoff of 200 miles following \citet{Conley_1999}. The column labeled `Diff-in-Diff' estimates (\ref{eq:tva}) by OLS and is similar to the estimate reported in \citet{Kline_Moretti_2014}. The final four columns labeled `Diff-in-Diff with Spillovers' are estimates from (\ref{eq:tva_spillover}).
                
                \item $^{*} p< 0.1$; $^{**} p < 0.05$; $^{***} p < 0.01$.
            \end{tablenotes}
        \end{threeparttable}
    \end{adjustbox}
\end{table}

Turning to the specification that includes spillovers, equation (\ref{eq:tva_spillover}), column (2) contains a point estimate for $\tau$ and columns (3)-(6) contain point estimates of the spillover effects $\delta_j$. For agricultural employment, the point estimates show there was a decline in agriculture employment in control units near the Authority. For control units between 0 and 50 miles, column (4) indicates a decline in agricultural employment of $3.7\%$ per decade. Between 50 and 100 miles the point estimate is $-1.6\%$ per decade, between 100 and 150 miles the point estimate is $-3\%$ per decade, and between 150 and 2000 miles the point estimate is $-1.6\%$ per decade. This is likely because higher-paying manufacturing jobs within the Authority drew farm-worker migrants from nearby counties. Because the spillovers onto the control counties are negative, the original difference-in-differences estimator was positively biased. The new point estimate indicates a decline of agricultural employment of about $7.4\%$ per decade compared to $5.1\%$ in the standard difference-in-difference specification. 

For manufacturing, our point estimates for spillovers are consistently negative, though imprecisely estimated in some columns. The spillover estimates suggest that neighboring counties experienced potentially negative spillover effects in the long run. Since there are negative spillover effects present, the new point estimate in column (2) of $3.5\%$ is significantly smaller than the original estimate of $5.6\%$. The spillover estimates are evidence that in the long run, urban shadow forces dominate the benefits of urban access. As you move away from the Tennessee Valley, the point estimates become more negative which suggests that the benefits of urban access vanish over distance faster than the costs of urban shadow.

To see how spillover effects from a large-scale place-based policy develop over time, Panel B in Table \ref{tab:tva} presents results for the effects of the Tennessee Valley in the short run using outcome data in 1960. Unlike in the long run, areas near the Tennessee Valley did not experience significant declines in agricultural employment in the short run. Since our long run analysis finds significant increases in high-paying manufacturing employment in the Tennessee Valley, this result is consistent with long run migration costs being lower than short run costs.

For manufacturing, there are potentially positive increases in manufacturing employment within 100 miles of the Tennessee Valley Authority and near-zero effects between 100 and 200 miles. In the short run, it appears that the effects of urban access and the cheap wholesale electricity dominated the effects of urban shadow. The effect of urban shadows can potentially be smaller in the short run if operating firms are unlikely to relocate. long run effects can be larger as entrant firms change their location decision and operating firms are slowly replaced which is what the long run spillover effect estimates suggest.

These results show that including spillovers in the estimation of treatment effects is potentially important and can lead to \emph{significant} differences in treatment effect estimates. Analysis of place-based policies that do not account for the fact that treatment effects can spill beyond the borders of treated areas can potentially be biased. More, the spillover effects caused by place-based policies change over time as frictions can create delays in reoptimizing behavior. 

\subsection{Identification Strategies in Place-Based Policy Analysis}\label{sec:pbp}

More generally, my framework provides important insights into identification strategies when analyzing the effects of place-based policies. There are two ways I contribute. First, \citet{Baum-Snow_Ferreira_2015} recognize the problem of spatial spillovers causing problems in identification and point to aggregation of units as a way to alleviate the problem (e.g. aggregating census tracts to metropolitan areas). However, this approach combines the treatment and spillover effects, which might each be of interest, into a singular aggregate effect. I have shown that all three effects of place-based policies can be estimated under very general assumptions about the spillovers. Therefore, the methods propose simple ways to estimate treatment effects with non-aggregated data.

Second, my framework provides insight into different identification strategies often used in the literature. Since place-based policies are targeted to specific distressed areas, comparison units are often hard to find. Researchers have developed many identification strategies to find comparison units that are similar in terms of unobservables. First, researchers use border discontinuities to compare treated units to units just on the other side of the border. Other times, researchers compare approved applicants to narrowly rejected applicants. Both strategies aim to balance unobservables between treated and control units. A key difference in these identification strategies is the distance comparison units are from the treated area and therefore the former is more prone to bias due to spillovers than the latter.

For example, consider the analysis of the effects of federal Empowerment Zones. The federal Empowerment Zones are specially designated areas in high-unemployment areas. The program gives businesses located in the zone tax incentives and the goal of the program is to reduce unemployment and poverty. In the literature, there are a set of conflicting results with some papers suggesting that the Empowerment Zones do indeed reduce poverty rates and others finding near-zero effects.\footnote{See Table 1 of \citet{Neumark_Young_2019} for a summary of the various treatment effect estimates in the literature.} 

\citet{Busso_Gregory_Kline_2013} compare census tracts in Empowerment Zones to census tracts that qualified and were rejected from the program. The rejected tracts are not typically geographically near accepted Empowerment Zones and they find large significant reductions in poverty rates. Meanwhile, \citet{Neumark_Kolko_2010} compare census tracts in Empowerment Zones to census tracts within 1,000 feet of the Zone. These control counties are likely the ones that experience the largest spillover effects and they find near-zero effects on poverty. 

My framework can reconcile both of these results. If census tracts just outside the Empowerment Zones also benefit from the policy, then the estimates of \citet{Neumark_Kolko_2010} are attenuated towards zero. These two identification strategies of using rejected applicants or using bordering units as the control group are very common in the urban literature more generally \citep{Baum-Snow_Ferreira_2015}. My paper suggests that the former is the preferred strategy if spillovers occur onto nearby control units. Identification strategies that rely on geographically close control units as having similar unobservables should be used cautiously. If treatment affects units on the other side of a border, then estimated treatment effects can be biased. In Appendix \ref{sec:oz}, I revisit analysis by \citet{Chen_Glaeser_Wessel_2021} on the 2017 Opportunity Zone program who find differing results between the `nearby' and `eligibe, but not selected' methodologies and show that my framework is able to reconcile their differences in estimates.

\section{Estimating Event Study Specifications with Spillovers}
\label{sec:event_study}

Now, I turn to the common setting where treatment turns on at different times for different units, often called staggered adoption. The intuition from how spillovers cause biases in estimates of the total effect of treatment on the treated extends into the staggered adoption setting. In this setting, the literature has shown that two-way fixed-effect estimates can be viewed as a weighted sum of $2 \times 2$ difference-in-differences estimates.\footnote{Various forms for these weights are described in \citet{Goodman-Bacon_2018}, \citet{Sun_Abraham_2020}, and \citet{deChaisemartin_DHaultfoeuille_2019}. I do not recharacterize the weights in this article and guide interested readers to the source articles themselves.} Therefore the bias terms will be an identically weighted sum of the bias term(s) from the $2 \times 2$ estimates, assuming that the parallel counterfactual trends assumption (\ref{assumption:parallel}) holds for all groups. However, since weights on some of the $2 \times 2$ estimates can be negative, the sign of the spillover effects does not determine the sign of the weighted average of the spillover effects. This makes the bias from spillovers much more difficult to sign. 

To estimate treatment and spillover effects in the presence of spillovers and staggered treatment timing, I will propose an estimation strategy that follows the `imputation-based' approach proposed in concurrent work by \citet{Gardner_2021} and \citet{Borusyak_Jaravel_Spiess_2021} by incorporating spillovers directly into estimation.\footnote{Our identification results in section \ref{sec:po_framework} lend themselves to similar identification results as in \citet{Callaway_SantAnna_2020}. One advantage of this is that covariates are flexibly included to allow for parallel-trends to hold conditionally on a set of covariates.} I assume a sample of independent and identically distributed panel for units $i$ and periods $t$ again taking the distribution of treatment and exposure treated as fixed. The imputation-based method relies on a model-based assumption for untreated and unexposed potential outcomes to formalize parallel trends:
\begin{assumption}[Parallel Counterfactual Trends (Staggered)]\label{assumption:parallel_staggered}
    For all units and all periods, the untreated and unexposed outcome is given by
    \begin{equation}\label{eq:y0_staggered}
        Y_{it}(0, \vec{0}) = \mu_i + \lambda_t + \varepsilon_{it},
    \end{equation}
    where $\expec{\varepsilon_{it}} = 0$.
\end{assumption}
This formalizes parallel trend by imposing that there is a common time-trend in the absence of treatment, $\lambda_t$. This is a particularly strong form of parallel trends in that it imposes units at every distance from treatment are on the same parallel trend $\lambda_t$ (see the discussion in section 2 of \citet{Borusyak_Jaravel_Spiess_2021}). Alternative parallel trends assumptions similar to assumption \ref{assumption:parallel-spill} can be used that assume parallel trends for all treatment timing-groups (units that start experiencing treatment in the same period) and spillover timing-groups (units that start having $S_{it} = 1$ in the same period) be equal to the never-affected control group. 

Note that assumption \ref{assumption:parallel_staggered} does not impose any structure on the effects of treatment and exposure $(D_{it}, h_i(\vec{D}_t))$ where now the exposure mapping is a function of the current period's treatment vector $\vec{D}_t \equiv \{D_{1t}, \dots, D_{nt} \}$. I also assume that units do not have any anticipatory effects before treatment starts:
\begin{assumption}[No Anticipation (Staggered)]\label{assumption:no-anticipation_staggered}
    For all observations with $D_{it} = 0$ and $h_i(\vec{D}_t) = 0$, $Y_{it}(D, \vec{h}) = Y_{it}(0,0)$ for all values of $D$ and $\vec{h}$.
\end{assumption}

With a model for $Y_{it}(0, \vec{0})$, treatment effects for individual $i$ and time $t$ under treatment status $D_{it}$ and exposure $h_i(\vec{D})$ are given by $\tau_{it}(D_{it}, h_i(\vec{D})) = Y_{it}(D_{it}, h_i(\vec{D})) - Y_{it}(0, \vec{0})$. Then, the total effect is formed similar to above:
\[
    \tau_{\text{total}} \equiv \expec{ \tau_{it}(1, h_i(\vec{D}_t)) \ \vert \ D_{it} = 1}
\]
However, it is also common in event-study analyses to allow heterogeneity in effects by the number of years that a unit experiences treatment. Let $K_{it}$ denote the number of years since treatment turns on (-1 for the year prior, 0 for the initial year, and so on). Then we can define dummy variables $D_{it}^k \equiv D_{i} \one_{K_{it} = k}$ and estimate average total effects for each relative period: 
\[
    \tau_{\text{total}}^k \equiv \expec{ \tau_{it}(1, h_i(\vec{D})) \ \vert \ D_{it}^k = 1}.
\]

Our identification argument follows from noting that under \nameref{assumption:parallel_staggered},
$$
  \expec{Y_{it}(D_{it}, h_i(\vec{D}_t)) - \mu_i - \lambda_t \ | \ \Omega} = \expec{\tau_{it}(D_{it}, h_i(\vec{D}_t)) \ | \ \Omega},
$$
where $\Omega$ is a set of observations $(i, t)$. Estimation of $\tau_{it}$ is the difference between the observed $Y_{it}$ and the unobserved outcome $Y_{it}(0, \vec{0})$. Similar to the previous section, under assumption (\ref{assumption:local}) we define $S_{it}$ to be an indicator equal to one if in period $t$, unit $i$ is within $\bar{d}$ distance from the nearest treated unit. 

Following the procedure laid out in \citet{Gardner_2021}, I propose a modified version of the two-stage difference-in-differences estimator: 
\begin{enumerate}
    \item Estimate $Y_{it} = \mu_i + \lambda_t + u_{it}$ for observations with $D_{it} = 0$ and $S_{it} = 0$ to estimate the common trend $\lambda_t$ and unit fixed-effects $\mu_i$. For all observations, residualize $\tilde{Y}_{it} \equiv Y_{it} - \hat{\mu}_i - \hat{\lambda}_t$ which is our estimate for $\tau_{it}$.
    
    \item Regress $\tilde{Y}_{it}$ on treatment and spillover dummy variables (discussed in table (\ref{tab:second_stage})).
\end{enumerate}

The first stage uses untreated and unexposed outcomes to estimate (\ref{eq:y0_staggered}). Then, averaging $\tilde{Y}_{it}$ in the second state will be unbiased and consistent for corresponding averages of $\tau_{it}$ under \nameref{assumption:parallel_staggered}. Table \ref{tab:second_stage} provides examples of treatment effects of interest and corresponding dummy variables to include.\footnote{Recent work by \citet{goldsmith2022contamination} suggests that multiple mutually exclusive treatment variables can create contamination bias. This is not a concern in this setting because the above imputation strategy avoids residualizing the treatment variables.}

\begin{table}[!tb]
    \caption{Second Stage Variables}
    \label{tab:second_stage}

    \centering
    \begin{threeparttable}
        \begin{tabular}{@{} *{2}{l} @{}}
            \toprule
            \textbf{Estimand} & \textbf{Included Variables} \\

            \midrule
            Total Effect & $D_{it}$ \\
            Total Effect (Event Study) & $D_{it}^k$ dummies \\
            Spillover Effect on Control & $S_{it}(1 - D_{it})$ or $\text{Ring}_{it}(1-D_{it})$ \\
            Spillover Effect on Control (Event Study) & $S_{it}^k(1 - D_{it})$ or $\text{Ring}_{it}^k(1-D_{it})$ \\
            \bottomrule
        \end{tabular}
    \end{threeparttable}
\end{table}

As discussed in \citet{Gardner_2021}, inference must account for the fact that the regressand in the second stage is estimated in the first stage. \citet{Gardner_2021} propose reframing the two-stage process as a two-stage GMM estimator and discusses how to perform valid inference in this setting. This procedure is implemented in the R/Stata package \texttt{did2s} \citep{did2s}. 

In Appendix \ref{sec:chc}, I use this two-stage estimator to revisit the analysis of \citet{Bailey_Goodman_Bacon_2015}. In particular, I find that the effects of community health centers stay within the county and therefore the originally estimated effects are robust. 

\section{Discussion}
\label{sec:conclusion}

This paper has considered the common environment where treatment is assigned via administrative boundary while the effects of treatment spread across these borders. I model this phenomenon in a potential outcomes framework and show why difference-in-differences estimation does not identify any treatment effect of interest. I then improve upon ad-hoc estimation strategies by begining with identification of well-defined treatment effects and then proposing estimation strategies that are robust to more forms of spillovers. In particular, I find that specifications with an indicator for being ``close to'' treated units will remove all bias in the treatment effect estimate so long as all units affected by spillovers are contained in the indicator under a modified parallel trends assumption. More, I show that a set of concentric `rings' can remove all the bias and improve the estimation of spillover effects. A limitation of this research is in deciding how wide and how many rings to include in estimation. Concurrent work by \citet{Butts_2021} discusses data-driven ring selection under a more stringent parallel trends assumption that does not readily apply in the context of large geographic units such as counties. 

Then, I show that in settings with spatial spillovers, estimates can change significantly. In particular, place-based interventions change the nature of agglomeration in the local and surrounding area--that is cause spillovers--local effects of these policies can be misestimated without controlling for general equilibrium effects. I also show the importance of considering spillovers in weighing the pros and cons of various identification strategies. Identification strategies based on geographic continuity of unobservables can magnify the bias from spillovers as they restrict the comparison group to observations experiencing the largest spillover effects.

\setlength{\bibsep}{0.0pt}
\bibliography{references.bib}

\appendix 
\numberwithin{equation}{section}
\numberwithin{figure}{section}
\numberwithin{table}{section}
\renewcommand{\thefigure}{\Alph{section}\arabic{figure}}
\renewcommand{\thetable}{\Alph{section}\arabic{table}}

\section{Proofs}
\label{sec:proofs}

\subsection{Proof of Proposition \ref{thm:bias}}

\begin{align*}
    &\underbrace{\expec{Y_{i1} - Y_{i0} \mid D_i = 1 } - \expec{Y_{i1} - Y_{i0} \mid D_i = 0 }}_{\text{Difference-in-Differences}} \\
    &= \expec{Y_{i1}(1, h_i(\vec{D})) - Y_{i0}(0, \vec{0})  \mid D_i = 1 } - \expec{Y_{i1}(0, h_i(\vec{D})) - Y_{i0}(0, \vec{0}) \mid D_i = 0 } \\
    &= \expec{Y_{i1}(1, h_i(\vec{D})) - Y_{i0}(0, \vec{0})  \mid D_i = 1 } \\
    &\quad\quad - \expec{Y_{i1}(0, h_i(\vec{D})) + Y_{i1}(0, \vec{0}) - Y_{i1}(0, \vec{0}) - Y_{i0}(0, \vec{0}) \mid D_i = 0 } \\
    &= \expec{Y_{i1}(1, h_i(\vec{D})) - Y_{i0}(0, \vec{0})  \mid D_i = 1 } - \expec{Y_{i1}(0, \vec{0}) - Y_{i0}(0, \vec{0}) \mid D_i = 0} \\ 
    &\quad\quad - \expec{Y_{i1}(0, h_i(\vec{D})) - Y_{i1}(0, \vec{0})\mid D_i = 0} \\ 
    &= \expec{Y_{i1}(1, h_i(\vec{D})) - Y_{i0}(0, \vec{0})  \mid D_i = 1 } - \expec{Y_{i1}(0, \vec{0}) - Y_{i0}(0, \vec{0}) \mid D_i = 1} \\
    &\quad\quad - \expec{Y_{i1}(0, h_i(\vec{D})) - Y_{i1}(0, \vec{0})\mid D_i = 0} \\  
    &= \expec{Y_{i1}(1, h_i(\vec{D})) - Y_{i0}(0, \vec{0}) - Y_{i1}(0, \vec{0}) + Y_{i0}(0, \vec{0})\mid D_i = 1 } \\
    &\quad\quad - \expec{Y_{i1}(0, h_i(\vec{D})) - Y_{i1}(0, \vec{0})\mid D_i = 0}\\
    &= \expec{Y_{i1}(1, h_i(\vec{D})) - Y_{i1}(0, \vec{0}) \mid D_i = 1 } - \expec{Y_{i1}(0, h_i(\vec{D})) - Y_{i1}(0, \vec{0})\mid D_i = 0} \\
    &\equiv \tau_{\text{total}} - \tau_{\text{spill}}(0), 
\end{align*}
where the fourth equality follows from our parallel trends assumption \ref{assumption:parallel} and the final equality follows from the definitions.

\subsection{Proof of Proposition \ref{thm:remove_bias}}

\begin{align*}
    &\expec{Y_{i1} - Y_{i0} \ \mid \ D_i = 1} - \expec{Y_{i1} - Y_{i0} \ \mid \ D_i = 0, S_i = 0} \\
    &\quad\quad= \expec{Y_{i1}(1, h_i(\vec{D})) - Y_{i0}(0,0) \ \mid \ D_i = 1} - \expec{Y_{i1}(0, 0) - Y_{i0}(0,0) \ \mid \ D_i = 0, S_i = 0} \\
    &\quad\quad= \expec{Y_{i1}(1, h_i(\vec{D})) - Y_{i0}(0,0) \ \mid \ D_i = 1} - \expec{Y_{i1}(0, 0) - Y_{i0}(0,0) \ \mid \ D_i = 1} \\
    &\quad\quad= \expec{Y_{i1}(1, h_i(\vec{D})) - Y_{i0}(0,0) - Y_{i1}(0, 0) + Y_{i0}(0,0) \ \mid \ D_i = 1} \\
    &\quad\quad= \expec{Y_{i1}(1, h_i(\vec{D})) - Y_{i1}(0, 0) \ \mid \ D_i = 1} \\
    &\quad\quad\equiv \tau_{\text{total}},
\end{align*}
where the first equality comes from the fact that $S_i = 0$ implies $h_i(\vec{D}) = 0$, the second equality from parallel trends assumption (\ref{assumption:parallel-mod}), and the last by definition of the total effect.

\subsection{Identification of Imputation Estimator}

This section will give identification arguments for the total effect and spillover effects. First, note under \nameref{assumption:parallel_staggered}, we have 
\begin{align*}
  \expec{Y_{it}(D_{it}, h_i(\vec{D}_t)) - \mu_i - \lambda_t \ | \ D_{it} = 1} 
  &= \expec{\tau_{it}(D_{it}, h_i(\vec{D}_t)) + \varepsilon_{it} \ | \ D_{it} = 1} \\
  &= \expec{\tau_{it}(D_{it}, h_i(\vec{D}_t)) \ | \ D_{it} = 1} \\
  &= \tau_{total}.
\end{align*}
In the first stage, we estimate $\mu_i$ and $\lambda_t$ using observations with $D_{it}$ and $S_{it} = 0$. Unbiasedness/consistency follows from \nameref{assumption:local} and \nameref{assumption:no-anticipation_staggered} and the arguments of the appendix of \citet{Gardner_2021}.

A similar argument follows for spillover effects. Again, assuming \nameref{assumption:parallel_staggered}, we have 
\begin{align*}
  \expec{Y_{it}(D_{it}, h_i(\vec{D}_t)) - \mu_i - \lambda_t \ | \ D_{it} = 0, S_{it} = 1} 
  &= \expec{\tau_{it}(D_{it}, h_i(\vec{D}_t)) + \varepsilon_{it} \ | \ D_{it} = 0, S_{it} = 0} \\
  &= \expec{\tau_{it}(D_{it}, h_i(\vec{D}_t)) \ | \ D_{it} = 0, S_{it} = 1}.
\end{align*}

\section{Application on Opportunity Zones}\label{sec:oz}

In this section, I revist the analysis of \citet{Chen_Glaeser_Wessel_2021} on the 2017 Opportunity Zone program which created tax incentives for captial investment in targeted Census Tracts.\footnote{Concurrent work by \citet{Arefeva_2021} also estimates spillover to nearby opportunity zones using a different datset and finds positive spillover effects on nearby census tracts as well.} To measure the affects of the opportunity zone program on housing prices, the authors collect a panel of census tracts with a measure of housing prices from the Federal Housing Finance Agency (FHFA) from 2014-2019.\footnote{The index tries to create a consistent price index that captures for changes in the composition of homes that are sold over time.} They produce estimates using two different identification strategies.

First, they compare census tracts that were selected as Opportunity Zones to eligible, but ultimately not selected, census tracts. As discussed in Section \ref{sec:pbp}, this estimation strategy relies on the assumption that since these census tracts are similar in nature to the treated census tracts (both meeting the program's criteria), it is likely that home prices would continue on similar trajectories in the absence of the program. The authors run a standard two-way fixed effect specification on the subsample of eligible census tracts:
\begin{equation}\label{eq:oz-eligible}
    Y_{it} = \mu_i + \mu_t + \tau D_{it} + \varepsilon_{it},
\end{equation}
where $Y_{it}$ is the annual change in the home price index, $\mu_i$ are tract fixed-effects, $\mu_t$ are time fixed effects, and $D_{it}$ is an indicator for treatment. 

The second identification strategy relies on comparing selected census tracts to geographically neighboring census tracts. As discussed in Section \ref{sec:pbp}, this estimation strategy relies on the assumption that proximity of census tracts would face similar economic shocks and hence home prices would likely evolve in parallel. For each treated census tract, they find the nearest non-treated census tract to form a pair, $(i, \tilde{i}) \equiv \nu$. They then estimate the following equation:
\begin{equation}\label{eq:oz-neighbor}
    Y_{it} - Y_{\tilde{i}t}  = \mu_\nu + \tau D_{it} + u_{it},
\end{equation}
where $Y_{it}$ is the annual change in the home price index, $\mu_\nu$ are pair fixed-effects and $D_{it}$ is an indicator for treatment. 

In both cases, the authors provide evidence of parallel `pre-trends' as evidence supporting the respective underlying assumptions. The results of both estimation strategies are shown in Table \ref{tab:oz}. Column (1) and (2) show the results of estimating Equation (\ref{eq:oz-eligible}) and (\ref{eq:oz-neighbor}) respectively. The `not-selected' estimate finds a marginally significant effect of an increase in home prices of about 0.3\% annually while the `neighboring' estimate finds a strongly significant effect twice as large of about 0.65\%.\footnote{The authors use this estimate to rule out effects larger than $\approx 0.65 + 2 * 0.25 = 1.15\%$. As shown before, these estimates are biased upwards and hence the upper bound of effect size can be lowered to about $\approx 0.65\%$.} 

\begin{table}[!tb]
    \caption{Effects of Opportunity Zones on Annual Home Price Growth}
    \label{tab:oz}
    \renewcommand{\arraystretch}{1.1}

    \begin{adjustbox}{width = 0.8\textwidth, center}
        \begin{threeparttable}
            \begin{tabular}{@{} l ccc @{}}
                \toprule

                & \multicolumn{1}{c}{\textbf{Not-Selected}} & \multicolumn{1}{c}{\textbf{Neighboring}} & \multicolumn{1}{c}{\textbf{Not-Selected and Neighboring}} \\ 
                \textit{Dependent Var.} & (1) & (2) & (3) \\

                \midrule
                
                      Treat $\times$ Post         & 0.3033$^{*}$    & 0.6478$^{***}$              & 0.1788\\   
                            & (0.1661)        & (0.2457)                    & (0.1692)\\   
       < 1/2mi. $\times$ Post      &                 &                             & -1.057$^{***}$\\   
                            &                 &                             & (0.3618)\\   
       < 1mi. $\times$ Post        &                 &                             & -0.7430$^{***}$\\   
                            &                 &                             & (0.1922)\\

                \bottomrule
            \end{tabular}
            
            \begin{tablenotes}\footnotesize
                \item \textit{Notes.} This table contains estimation 
                \item $^{*} p< 0.1$; $^{**} p < 0.05$; $^{***} p < 0.01$.
            \end{tablenotes}
        \end{threeparttable}
    \end{adjustbox}
\end{table}

What is driving the differences in these estimates? As I proposed in section \ref{sec:pbp}, the differences could be due to the fact that neighboring units could be experiencing effects from the Opportunity Zones. To test this, I use the estimation strategy proposed in section \ref{sec:estimation} to modify equation \ref{eq:oz-eligible}. For the subsample of eligible census tracts, I run the following specification: 
\begin{equation}\label{eq:oz-spill}
    Y_{it} = \mu_i + \mu_t + \tau D_{it} + \gamma_{1} \text{Within 1/2mi.} +  \gamma_{2} \text{Within 1mi.} + \varepsilon_{it},
\end{equation}
where $\text{Within}$ are indicators for being within 1/2mi. and being between 1/2 and 1mi. from an Opportunity Zone. This estimation strategy uses non-neighboring census tracts as the effective control group and estimates effects for census tracts within or close to Opportunity Zones. The results are estimated in Table \ref{tab:oz} in Column (3). These estimated treatment effect decreased in this specification to 0.17\% and reveals that census tracts just outside of the Opportunity Zone experience negative and significant declines in home prices of about 1\%. This result explains why the estimated effect using the `neighbor' speficiation is twice as large as the `eligible' specification. 

The results of this section highlight the importance of considering spillovers when using "neighbor" style identification strategies, especially in settings where the unit of observation is small and spillover effects are quite likely. 

\section{Application on Community Health Centers}\label{sec:chc}

As an application of the above methods, I extend the analysis of \citet{Bailey_Goodman_Bacon_2015}. The authors study the creation of federal community health centers between 1965 and 1974 that provided \textit{primary} care to low-income communities. They test the hypothesis if access to low-/no-cost health care services decreased mortality rates on the treated population. To answer this question, the authors use a common event-study framework to compare outcomes in treated counties to all other US counties by estimating the following specification 
\begin{equation}\label{eq:chc_es}
    Y_{it} = \theta_i + X_{it} \beta + \sum_{k = -7}^{-2} \pi_y D_{it}^k + \sum_{k = 0}^{15} \tau_{y} D_{it}^k + \varepsilon_{it},
\end{equation}
where $D_i^k$ is the typical `event study' indicator for being treated for $k$ years, $\theta_i$ is county fixed-effects, and $X_{it}$ contains a set of controls.\footnote{Controls include 1960 county characteristic trends, state-year fixed-effects, and urban-group fixed-effects. A full list can be found on page 1080 of \citet{Bailey_Goodman_Bacon_2015}.} The coefficents $\pi_y$ can be interpreted as tests of parallel pre-trends and $\tau_y$ can be interepreted as the treatment effect of a community health center $y$ years after establishment. In years following the establishment of the community health centers, the authors find a reduction of between 15-30 deaths per 100,000 residents compared to a baseline adjusted mortality rate of 929 deaths per 100,000 residents. 

\begin{figure}[tb!]
    \caption{Total and Spillover Effects of Community Health Centers}
    \label{fig:chc_es_spill}
        
    {\centering
        \includegraphics[width=\textwidth]{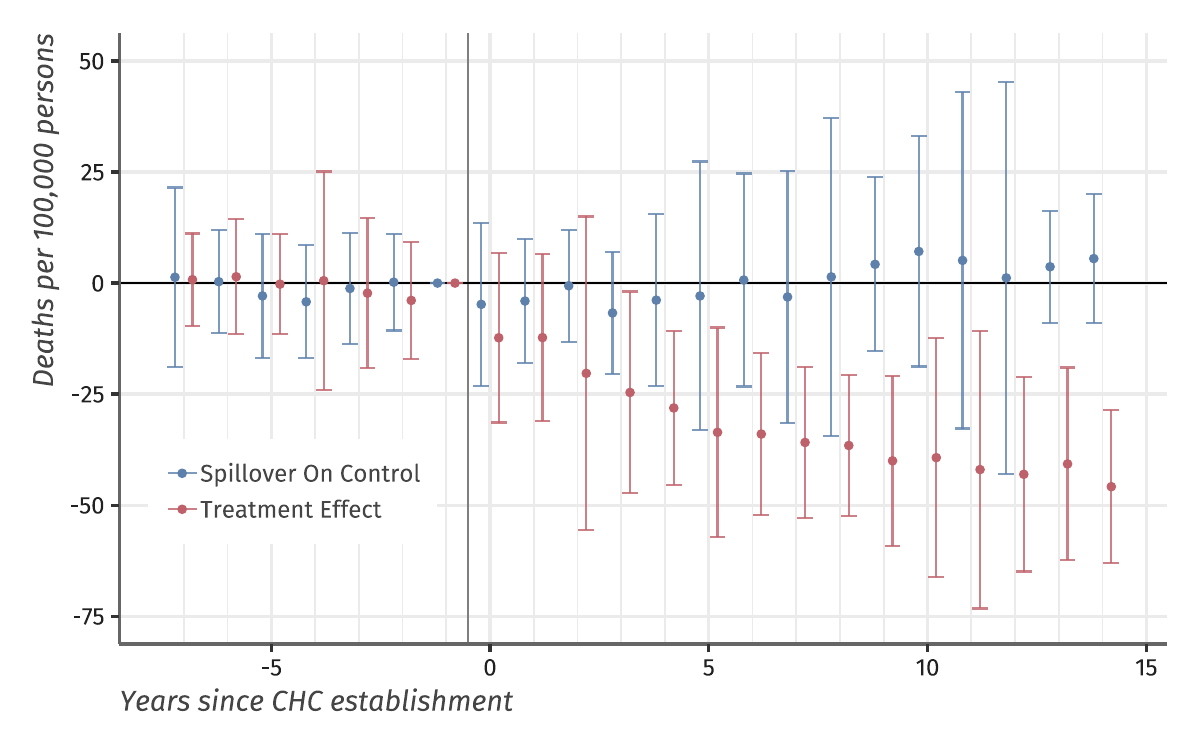}
    }
    {\footnotesize
        \textit{Notes:} This figure plots event study estimates for the total effect and the spillover effect on control units within 25 miles of treatment at different periods relative to establishment year. The estimates are generated using the `did2s' package \citep{did2s}. 
    }
\end{figure}

There are theoretical reasons to think spillovers may or may not exist in this context. On the one hand, individuals outside the county can potentially travel to the community health centers to receive care. This would create a negative spillover effect on mortality rates in nearby counties which would bias their estimates towards zero. On the other hand, \citet{Bailey_Goodman_Bacon_2015} document evidence that their estimated effects are not due to emergencies but rather primary care services. In this case, it is less likely low-income individuals would travel very far to receive primary care, and hence the spillover effects are potentially near zero. My methodology can provide an answer to the question of how far do individuals travel for low-/no-cost primary care.

As in the method detailed above, I use an indicator for being within 25 miles of a treated county as the spillover variable. The results are presented in Figure \ref{fig:chc_es_spill}. The confidence intervals labeled with circles represent point estimates for the average spillover effect on control units within 25 miles. No spillover effect is estimated to be significantly different from zero which suggests that the effects of community health centers are very local. Since there are near zero spillover effects, the total effect estimates marked in Figure \ref{fig:chc_es_spill} as diamonds maintain the same shape as the author's original estimates with estimates between 15-30 fewer deaths per 100,000 persons. 

The spillover effects results provide evidence that low-income individuals will not travel far to receive primary care. Practically, this suggests that community health centers should be targeted to be as accessible as possible for poor individuals as they are unable to travel far to access the services. 

\end{document}